\RequirePackage{fix-cm}
\documentclass[smallextended]{svjour3}       
\smartqed  
\usepackage[pdftex]{graphicx}
\usepackage{braket}
\usepackage{bm}
\usepackage{lscape}
\usepackage{amsmath}
\usepackage{amssymb}
\usepackage{comment}
\usepackage[root-radius=0.15cm,edge length=0.8cm,mark=o]{dynkin-diagrams}
\newcommand{\Kcbulk}{K_{\mathrm{c}}^{\mathrm{bulk}}}
\newcommand{\Ks}{K_{\mathrm{s}}}
\newcommand{\hs}{h_{\mathrm{s}}}
\newcommand{\Dcbulk}{D_{\mathrm{c}}^{\mathrm{bulk}}}

\newcommand{\eq}[1]{Eq.~(\ref{#1})}
\journalname{Journal of Statistical Physics}
\begin{document}

\title{Boundary CFT and tensor network approach to surface critical phenomena of the tricritical 3-state Potts model
}

\titlerunning{Surface critical phenomena of the tricritical 3-state Potts model}

\author{Shumpei Iino}


\institute{S. Iino \at
  Institute for Solid State Physics, The University of Tokyo,
  Kashiwa, Chiba, Japan \\
  \email{iino@issp.u-tokyo.ac.jp}           
}


\maketitle

\begin{abstract}
  One-dimensional edges of classical systems in two dimension sometimes show surprisingly rich phase transitions and critical phenomena, particularly when the bulk is at criticality. As such a model, we study the surface critical behavior of the 3-state dilute Potts model whose bulk is tuned at the tricritical point. To investigate it more precisely than in the previous works~[Y. Deng and H. W. J. Bl\"ote, Phys. Rev. E \textbf{70}, 035107(R) (2004); Phys. Rev. E \textbf{71}, 026109 (2005)], we analyze it from the viewpoint of the boundary conformal field theory (BCFT). The complete classification of the conformal boundary conditions for the minimal BCFTs discussed in [R. E. Behrend et al., Nucl. Phys. B \textbf{579}, 707 (2000)] allows us to collect the twelve boundary fixed points in the tricritical 3-state Potts BCFT. Employing the tensor network renormalization method, we numerically study the surface phase diagram of the tricritical 3-state Potts model in detail, and reveal that the eleven boundary fixed points among the twelve can be realized on the lattice by controlling the external field and coupling strength at the boundary. The last unfound fixed point would be out of the physically sound region in the parameter space, similarly to the `new' boundary condition in the 3-state Potts BCFT.
\end{abstract}

\newpage
\section{Introduction}\label{sec:intro}

Surface critical behaviors are the critical phenomena that boundaries of the system exhibit, where the various physical quantities at the boundaries show singular behaviors with the critical exponents different from the bulk ones~\cite{Binder1983_DG}. Interestingly the universality classes of surface critical behaviors are much richer than for the ordinary bulk critical phenomena, since the critical behavior of the surfaces changes in general depending on the various boundary conditions (b.c.'s) imposed. For instance, suppose the three-dimensional classical Ising model with the ferromagnetic nearest-neighbouring interactions in the presence of open boundaries. The surface critical behavior with the disordered (or free) b.c. imposed is different from that with the ordered b.c. with the non-zero magnetization.

In this paper, we focus on the surface critical phenomena of the two-dimensional classical systems, which indicates the surfaces of the systems are one-dimensional classical edges. While no phase transition at the finite temperature occurs in the classical one-dimensional systems, nontrivial edge phenomena may happen in the presence of strong correlation in the two-dimensional bulk. An example is the edge phase transition of the Blume-Capel model with spin 1 at the tricritical point~\cite{Blume1966,Capel1966}, whose Hamiltonian is
\begin{equation}
  \beta\mathcal{H} = -\Kcbulk\sum_{\langle ij\rangle\in\mathrm{bulk}}\sigma_i\sigma_j
  +\Dcbulk\sum_{i}{\sigma_i}^2-\Ks\sum_{\langle ij\rangle\in\mathrm{edge}}\sigma_i\sigma_j
  -\hs\sum_{i\in\mathrm{edge}}\sigma_i,
  \label{eq:S1BC}
\end{equation}
with the inverse temperature $\beta$ and the Ising spin $\sigma_i=\pm 1$ or $0$ living in the $i$-th site. The coupling constant $\Kcbulk$ and the chemical potential $\Dcbulk$ in the bulk are tuned at the tricritical point, whose universality class is the tricritical Ising one. By controlling the surface coupling strength $\Ks$ and the surface external field $\hs$, various surface phase transitions can be observed~\cite{Affleck2000,Deng2004,Deng2005}. One remarkable feature of the surface critical behaviors in \eq{eq:S1BC} is that the edge can be ordered with finite surface coupling $\Ks$ even under $\hs=0$, which surprisingly implies the existence of `the finite temperature transition of the one-dimensional edge.' Notice that such a phenomenon can be physically allowed since the edge in this system is not the isolated one-dimensional system but the one connected to the strongly correlated two-dimensional bulk at the tricriticality. Similar surface transitions can be observed in higher dimensions such as the three-dimensional classical $O(3)$ and $O(4)$ models~\cite{Krech2000,Deng2005_2,Deng2006} and the two-dimensional quantum Heisenberg antiferromagnet~\cite{Zhang2017,Ding2018,Weber2018}.

The surface phase transitions with a critical bulk can be described by the boundary conformal field theory (boundary CFT; BCFT), which provides much information on the surface critical phenomena especially in two dimension, such as the possible boundary states invariant under the conformal transformations and the critical exponents of the surface transitions~\cite{Belavin1984,Cardy1984,Cardy1987_DG,Cardy2006}. The permissible boundary states in BCFT must satisfy the condition $T=\bar{T}$ at the boundary, where $T$ is the energy-momentum operator and $\bar{T}$ is the antiholomorphic counterpart. This restriction can be intuitively interpreted as absence of the energy flow across the boundaries. Although it has been demonstrated that the boundary states satisfying this restriction can be constructed for some specific cases~\cite{Behrend2000,Janik2001,Cappelli2002,Quella2002,Blakeley2009}, the complete classification of the conformal b.c.'s for the general BCFT is still an open problem.

The surface critical behavior of the tricritical Ising model referred to above is well studied in terms of both BCFT and the lattice model. Chim had constructed the conformal boundary states of the tricritical Ising BCFT~\cite{Chim1996}, and the surface phase diagram of this model was conjectured by Affleck by means of the BCFT arguments~\cite{Affleck2000}. Then, the consistency of the conjectured critical exponents with the lattice model \eq{eq:S1BC} was demonstrated by Deng and Bl{\"o}te, where they made use of the Monte Carlo (MC) simulation~\cite{Deng2004,Deng2005}.

In Deng and Bl{\"o}te's papers, besides the ordinary Blume-Capel model \eq{eq:S1BC}, they also investigated the surface phase diagram of the 3-state dilute Potts model, the generalized Blume-Capel model with the higher symmetry $S_3$ rather than $Z_2$:
\begin{align}
  \begin{split}
    \beta\mathcal{H} = &-\Kcbulk\sum_{\langle ij\rangle\in\mathrm{bulk}}\delta_{\sigma_i\sigma_j}(1-\delta_{\sigma_i0})
    -\Dcbulk\sum_{i}\delta_{\sigma_i0}\\
    &-K_{\mathrm{s}}\sum_{\langle ij\rangle\in\mathrm{edge}}\delta_{\sigma_i\sigma_j}(1-\delta_{\sigma_i0})
    -h_{\mathrm{s}}\sum_{i\in\mathrm{edge}}\delta_{\sigma_iA},
  \end{split}
  \label{eq:S3BC}
\end{align}
where $\delta$ is the Kronecker's delta, and each spin can take four different values, $\sigma=A,B,C,$ or $0$ ($0$ represents the vacancy). Notice that in the absence of the external field $\hs=0$ the Hamiltonian \eq{eq:S3BC} is invariant under the arbitrary permutation within the three spins $A,B,$ and $C$, and this model exhibits the tricritical 3-state Potts (TC3P) point when the bulk parameters are fine-tuned. Their MC simulation demonstrated that, similarly to the tricritical Ising case, the surface phase transition occurs at the finite surface coupling with $\hs=0$, and also at the finite surface magnetic field with $\Ks=0$.

However, the detailed analysis of the TC3P model from the viewpoint of the BCFT is still missing. The critical exponents and the phase diagram obtained by the MC simulation have not been checked precisely in terms of BCFT. Therefore the purpose of this paper is to make a list of the possible boundary fixed points in the TC3P BCFT and compare them with the numerically studied surface phase diagram of the Hamiltonian \eq{eq:S3BC}. Fortunately, the conformal boundary states in the minimal CFTs are completely classified by Behrend, Pearce, Petkova, and Zuber~\cite{Behrend1998_2,Behrend1998,Behrend2000}. Applying their formula to the TC3P BCFT allows us to collect the twelve conformal boundary fixed points together with some of their properties such as the operator contents and the Affleck and Ludwig's $g$-values~\cite{Affleck1991}.

For the obtained conformal boundary fixed points, we identify them in the surface phase diagram of the lattice TC3P model with the help of the conformal spectrum numerically extracted from the Hamiltonian \eq{eq:S3BC}. To compute the conformal spectrum from the lattice Hamiltonian, we perform the numerical simulation with the tensor network renormalization (TNR) method, by which we can obtain accurate scaling dimensions through the diagonalization of the transfer matrix~\cite{Evenbly2015}.

The remaining parts of this paper are devoted as follows. In Sec.~\ref{sec:BCFT}, we review the known results of the TC3P BCFT, where the minimally brief introduction to the BCFT is also presented. For the TC3P model, all the possible conformal boundary states are gathered using the $A$-$D$-$E$ classification of the minimal-series BCFTs, for each of which the operator content of the corresponding boundary fixed point and the $g$-value are calculated analytically. Next, we turn to the numerical approach to the lattice model in Sec.~\ref{sec:lattice}, where it is briefly explained how to compute the conformal spectrum from the lattice. We not only determine the accurate bulk tricritical point of \eq{eq:S3BC}, but also describe the surface phase diagram and discuss the correspondence to the boundary states obtained in Sec.~\ref{sec:BCFT}. Finally, we conclude our results in Sec.~\ref{sec:conclusion}.

\section{Boundary fixed points in the TC3P BCFT}\label{sec:BCFT}

In this section, we review what is known about the BCFT of the TC3P model. First we present a minimal introduction to the two-dimensional BCFT, which can be skipped if the reader is familiar with it. After a review on the modular invariant torus partition function of the TC3P model, we apply the formulae of the $A$-$D$-$E$ classification provided in Ref.~\cite{Behrend2000} to this model for the purpose of obtaining the full set of the conformal boundary states. This $A$-$D$-$E$ classification also allows us to calculate the $g$-values and the operator contents of the corresponding boundary fixed points, which are useful to study the surface phase diagram of the lattice model later.

\subsection{A minimal review of CFT and BCFT in two dimension}

For readers who are not familiar with CFT or BCFT, we briefly introduce some concepts necessary to understand the discussion in this paper. As for more detailed introduction, see textbooks such as Ref.~\cite{Ginsparg1989,Francesco_CFT,Henkel_CFT}.

Every CFT in two dimension, in general, can be characterized by a set of quantities called \textit{conformal data}: conformal weights of the primary operators, coefficients of the operator product expansions (OPEs) between those operators, and a scalar value $c$ called central charge~\cite{Belavin1984}. In the language of the representation theory of the Lie algebra, each primary operator corresponds to a highest weight state in an irreducible module whose highest weight is the conformal weight. The conformal algebra for the CFT defined on an infinite plane is the direct product of the two copies of the \textit{Virasoro algebra} (or its extension)~\cite{Virasoro1970}, whose structure constants between the infinite number of generators $L_n$ and $\bar{L}_n$ are given as
\begin{align}
  \begin{split}
    \left[L_m,L_n\right]&=(m-n)L_{m+n}+\frac{c}{12}m(m^2-1)\delta_{m+n,0}\\
    \left[\bar{L}_m,\bar{L}_n\right]&=(m-n)\bar{L}_{m+n}+\frac{c}{12}m(m^2-1)\delta_{m+n,0}\\
    \left[L_m,\bar{L}_n\right]&=0,
  \end{split}
  \label{eq:VirasoroAlgebra}
\end{align}
with integers $m$ and $n$. In fact, \eq{eq:VirasoroAlgebra} suggests that the algebra can be separable into the direct product of the holomorphic and the antiholomorphic part as $\{L_m\}\otimes\{\bar{L}_n\}$. Therefore, in a given CFT, when one describes the set of the primary operators of the holomorphic part as $\{\phi_i\}$, the direct products of two copies of the primary operators $\{\phi_i\}\otimes\{\bar{\phi}_j\}$ can occur in the theory.

For statistical systems defined on a torus, however, these decoupled degrees of freedom are often restricted, due to the modular invariance~\cite{Cardy1986_2}. Suppose the CFT on a torus geometry with the modular parameter $\tau$. In general, the partition function can be described as
\begin{equation}
  Z(\tau) = \sum_{i,j}\chi_i(q)N_{ij}\bar{\chi}_j(\bar{q}),
\end{equation}
where $q=\exp(2\pi i\tau)$ and $\chi_i(q)$ is the character of the irreducible module $i$ with $\bar{\chi}_j(\bar{q})$ being the antiholomorphic counterpart. The indices $i$ and $j$ run within all the possible irreducible modules respectively, and $N_{ij}$ is a non-negative integer which determines how many times the primary operators $\phi_i\otimes\bar{\phi}_j$ occur in the theory. In many cases one encounters in statistical mechanics including the TC3P model, the partition functions possess the modular invariance, which results in a strong restriction on $N_{ij}$.

The unitary minimal model is a special series of the two-dimensional CFTs, whose properties are well studied~\cite{Belavin1984}. Each of them can be characterized by two successive integers as $\mathcal{M}_{m+1,m}$, whose conformal data are completely determined. The modular invariant partition functions of this series can be characterized by a pair of the simply-laced simple Lie algebra, which is called $A$-$D$-$E$ classification~\cite{Cappelli1987}. In fact, the partition functions of many critical systems with discrete symmetries, such as Ising model and its multicritical versions, can be described by the modular invariants of the unitary minimal models~\cite{Friedan1984}.

Next, we turn to the two-dimensional CFT in the presence of open boundaries, which is often called BCFT~\cite{Cardy1984}. In this paper, we focus on the BCFT defined on a finite cylinder, where the b.c.'s labeled as $\alpha$ and $\beta$ are imposed on either side, respectively. These b.c.'s, as is referred to in Sec.~\ref{sec:intro}, must satisfy some restriction so that there should be neither inflow nor outflow of the energy across the boundaries. Cardy first proposed a general way of constructing the solution to this restriction as the linear superposition of a certain basis called \textit{Ishibashi states}~\cite{CARDY1989581,Ishibashi1989}:
\begin{equation}
  \ket{\alpha} = \sum_j\frac{S_{\alpha j}}{\sqrt{S_{1j}}}\ket{j},
  \label{eq:Ishibashi_def}
\end{equation}
where $\ket{\alpha}$ is the conformal boundary state corresponding to the b.c. $\alpha$, $\ket{j}$ is the Ishibashi state labeled by the irreducible module $j$, and the summation is taken for all the irreducible modules in the theory. $S$ is the so-called modular-$S$ matrix determining how the characters transform under the modular transformation $\tau\rightarrow -1/\tau$, whose explicit form is discussed later in this section. Notice that the index $1$ corresponds to the identity operator, and the overlap between two Ishibashi states are given as
\begin{equation}
  \braket{i|q^{\frac{1}{2}(L_0+\bar{L}_0-\frac{c}{12})}|j}=\delta_{ij}\chi_i(q).
\end{equation}
The conformal boundary states for the minimal models can also be completely classified in the similar form to \eq{eq:Ishibashi_def}~\cite{Behrend2000}.

Similarly to the case of the torus geometry, the partition function on the finite cylinder can be given as
\begin{equation}
  Z_{\alpha|\beta} = \sum_in^i_{\alpha\beta}\chi_i,
  \label{eq:Zab_general}
\end{equation}
where the coefficient $n^i_{\alpha\beta}$ is a non-negative integer, and $\chi_i$ is again the character of the irreducible module $i$~\cite{CARDY1986200}. Therefore, $n^i_{\alpha\alpha}$ determines the operator content of the boundary fixed point labeled by $\alpha$. Notice that since there is only a single copy of the Virasoro algebra acting on the space of states, the BCFT partition function \eq{eq:Zab_general} is a linear combination of characters, not a bilinear combination. As is discussed later, for the minimal series of the BCFT, the non-negative integer $n^i_{\alpha\beta}$ can be calculated explicitly based on the $A$-$D$-$E$ classification.

\subsection{The modular invariant partition function}\label{subsec:Z_of_TC3P}

While the ordinary unitary minimal CFT $\mathcal{M}_{7,6}$ with $c=6/7$ describes the critical phenomena of the pentacritical Ising model, the criticality of the TC3P model corresponds to the $c=6/7$ CFT with the extended symmetry of the conserved `parafermion currents'~\cite{Huse1984,Zamolodchikov1987}. There are fifteen primary fields in $\mathcal{M}_{7,6}$ specified by a pair of integers $(r,s)$ as shown in Tab.~\ref{tab:c=6/7}, only the nine of which appear in the TC3P CFT (the ones with $r=1,3,$ and $5$ in Tab.~\ref{tab:c=6/7})~\cite{Friedan1984,Itzykson1986_2,Zuber1986}. The torus partition function of the TC3P model is given by the nondiagonal $(D_4,A_6)$ modular invariant in terms of the $A$-$D$-$E$ classification, as
\begin{equation}
  Z_{D_4,A_6} = \sum_{s=1,2,3}\left[\left|\chi_{1,s}+\chi_{5,s}\right|^2+2\left|\chi_{3,s}\right|^2\right],
  \label{eq:Z_D4A6}
\end{equation}
where $\chi_{r,s}$ is the character for the irreducible module labeled by $(r,s)$ in the Virasoro algebra~\cite{Cappelli1987,Cappelli1987_2}.

Notice that defining the character for the extended algebra makes the partition function diagonal: 
\begin{equation}
  \begin{split}
    Z_{D_4,A_6} = \left|C_{0}\right|^2&+\left|C_{\frac{1}{7}}\right|^2+\left|C_{\frac{5}{7}}\right|^2
    +\left|C_{\frac{4}{3}}^+\right|^2+\left|C_{\frac{4}{3}}^-\right|^2\\
    &+\left|C_{\frac{10}{21}}^+\right|^2+\left|C_{\frac{10}{21}}^-\right|^2
    +\left|C_{\frac{1}{21}}^+\right|^2+\left|C_{\frac{1}{21}}^-\right|^2,
  \end{split}
  \label{eq:Z_D4A6_diag}
\end{equation}
where the new characters in the extended algebra are defined as
\begin{align}
  \label{eq:Z_D4A6_character_0}
  C_{0} &= \chi_{1,1}+\chi_{5,1} & C_{\frac{1}{7}} &= \chi_{1,2}+\chi_{5,2} & C_{\frac{5}{7}} &= \chi_{1,3}+\chi_{5,3},\\
  C_{\frac{4}{3}}^+ &= C_{\frac{4}{3}}^- = \chi_{3,1} & C_{\frac{10}{21}}^+ &= C_{\frac{10}{21}}^- = \chi_{3,2} & C_{\frac{1}{21}}^+ &= C_{\frac{1}{21}}^- = \chi_{3,3}.
  \label{eq:Z_D4A6_character_1}
\end{align}
As \eq{eq:Z_D4A6_diag}, (\ref{eq:Z_D4A6_character_0}), and (\ref{eq:Z_D4A6_character_1}) suggest, it is useful to reorganize the nine primary operators in the extended chiral algebra. While the three of them are invariant under the $Z_3$ transformations, $\phi_{0}$, $\phi_{\frac{1}{7}}$, and $\phi_{\frac{5}{7}}$, the others have the nontrivial $Z_3$ charge: let us define the operators with the positive $Z_3$ charge as $\phi_{\frac{4}{3}}^+$, $\phi_{\frac{10}{21}}^+$, and $\phi_{\frac{1}{21}}^+$, and for the negative charge $\phi_{\frac{4}{3}}^-$, $\phi_{\frac{10}{21}}^-$, and $\phi_{\frac{1}{21}}^-$.
\begingroup
\renewcommand{\arraystretch}{1.5}
\begin{table}
  \caption{The primary fields of the $c=6/7$ unitary minimal CFT. Note that the element at $(r,s)$ is identical to that at $(6-r,7-s)$.}\label{tab:c=6/7}
  \centering
  \begin{tabular}{|cc|cccccc|}\hline
     & 5 & $5$ & $\frac{22}{7}$ & $\frac{12}{7}$ & $\frac{5}{7}$ & $\frac{1}{7}$ & \\
     & 4 & $\frac{23}{8}$ & $\frac{85}{56}$ & $\frac{33}{56}$ & $\frac{5}{56}$ & & \\
    $r$ & 3 & $\frac{4}{3}$ & $\frac{10}{21}$ & $\frac{1}{21}$ & & & \\
     & 2 & $\frac{3}{8}$ & $\frac{1}{56}$ & & & & \\
     & 1 & $0$ & & & & & \\\hline
     &  & 1 & 2 & 3 & 4 & 5 & 6 \\
     &  & \multicolumn{6}{c|}{$s$} \\\hline
  \end{tabular}
\end{table}
\endgroup

\subsection{Conformal boundary states in the TC3P BCFT}\label{subsec:boundaryStates}

For the minimal CFTs characterized by a pair of the Dynkin diagrams of the simply-laced simple Lie algebra as $(G,A_{h-1})$ with $h$ being the Coxeter number, the permissible conformal boundary states in the corresponding BCFT are completely classified~\cite{Behrend2000}. In particular as for the $D$-type theories, in which the TC3P model is included, more detailed study on the OPE coefficients for the bulk and boundary fields can be found in Ref.~\cite{Runkel2000}. In this section, we calculate the conformal boundary states in the TC3P model employing the complete classification presented in Ref.~\cite{Behrend2000}.

According to the complete classification, the conformal boundary states can be characterized by a pair of labels as $\ket{(s,a)}$, where $s$ is an integer satisfying $1\leq s\leq h-1$ and $a$ specifies the nodes of the Dynkin diagram $G$. Notice that the notation of $r$ and $s$ in this paper is opposite in Ref.~\cite{Behrend2000}. Since the primary operator labeled by $(r,s)$ is identical to the one by $(g-r,h-s)$ with $g$ being the Coxeter number of $G$, the number of the conformal boundary states is $n(h-1)/2$, where $n$ is the number of nodes in $G$. In the case of the TC3P CFT with $g=6$, $h=7$, and $G=D_4$, there are $4\times (7-1)/2=12$ independent conformal boundary states.

The explicit forms of the conformal boundary states can be given as the linear combination of the Ishibashi states like \eq{eq:Ishibashi_def}:
\begin{equation}
  \ket{(s,a)} = \frac{\sqrt[4]{2}}{2}\sum_{\substack{r'\in\mathrm{Exp}(G)\\ 1\leq s'\leq h-1}}\frac{\psi_a^{r'}S^{(h)}_{ss'}}{\sqrt{S_{1r'}^{(g)}S_{1s'}^{(h)}}}\ket{r',s'},
  \label{eq:boundaryStateGeneralForm}
\end{equation}
where $\psi$ is the eigenvectors of the adjacency matrix for the Dynkin diagram $G$ and $\mathrm{Exp}(G)$ is the set of the Coxeter exponents of $G$, the number of whose elements is equal to $n$. $\ket{r',s'}$ is the Ishibashi state labeled by the primary operators, i.e., a pair of integers, as in Tab.~\ref{tab:c=6/7}. Note that $\ket{r',s'}$ is identical to $\ket{g-r',h-s'}$ just as the primary operators, which results in the prefactor $1/2$ in \eq{eq:boundaryStateGeneralForm}. $S^{(h)}$ is also the eigenvectors of the adjacency matrix of $A_{h-1}$, whose explicit form is
\begin{equation}
  S_{ij}^{(h)} = \sqrt{\frac{2}{h}}\sin\frac{ij\pi}{h},
  \label{eq:modularSh}
\end{equation}
with $1\leq i,j\leq h-1$.

Let us apply the formula \eq{eq:boundaryStateGeneralForm} to our interest, the TC3P BCFT. The way of calculating the matrix $\psi$ is explained in Appendix B of Ref.~\cite{Behrend2000}. First, we begin with naming the nodes of the Dynkin diagram $D_4$ as
\begin{equation}
  \dynkin[labels={1,2,(3,-),(3,+)}]D4.
  \label{eq:D4node}
\end{equation}
Although the Coxeter exponents of $D_4$ are $\{1,3,3,5\}$, we here discriminate the degenerated $3$'s as $(3,-)$ and $(3,+)$ similarly to the nodes of $D_4$ in \eq{eq:D4node}, which leads to $\mathrm{Exp}(D_4)=\{1,(3,-),(3,+),5\}$. Now the explicit form of the matrix $\psi$ can be given as follows:
\begin{align}
  \psi^r_a &= \sqrt{2}S_{ar}^{(g)}=\sqrt{\frac{2}{3}}\sin\frac{ra}{6}\pi & r,a&\neq 3\\
  \psi^{(3,\pm)}_a &= S_{a3}^{(g)}=\frac{\delta_{a,1}}{\sqrt{3}} & a&= 1,2\\
  \psi^r_{(3,\pm)} &= \frac{1}{\sqrt{2}}S_{3r}^{(g)}=\frac{1}{\sqrt{6}} & r&= 1,5\\
  \psi^{(3,\epsilon')}_{(3,\epsilon)} &= \frac{1}{2}\left(S_{33}^{(g)}+i\epsilon\epsilon'\right)=\frac{\omega^{\epsilon\epsilon'}}{\sqrt{3}} & \epsilon,\epsilon'&=\pm 1,
\end{align}
where $\omega = \exp\left(2\pi i/3\right)$, $r\in\mathrm{Exp}(D_4)$, and $a$ runs within the nodes of $D_4$ shown in \eq{eq:D4node}.

Using the above results, we can calculate the explicit forms of the conformal boundary states spanned by the Ishibashi states as follows:
\begin{align}
  \label{eq:symBroken1}
  \ket{(s,1)} &= \ket{(7-s,1)} 
  =\sum_{s'=1}^3P_{ss'}\left(\ket{1,s'}+\ket{5,s'}+\ket{3-,s'}+\ket{3+,s'}\right)\\
  \ket{(s,3\pm)} &= \ket{(7-s,3\pm)} 
  =\sum_{s'=1}^3P_{ss'}\left(\ket{1,s'}+\ket{5,s'}+\omega^{\mp}\ket{3-,s'}+\omega^{\pm}\ket{3+,s'}\right),
  \label{eq:symBrokenPM}
\end{align}
where $s=1,2,3$ and
\begin{equation}
  P_{ss'} = \left(
    \begin{array}{ccc}
      x & y & z \\
      \frac{y^2}{x} & -\frac{z^2}{y} & \frac{x^2}{z} \\
      \frac{z^2}{x} & \frac{x^2}{y} & -\frac{y^2}{z}
    \end{array}
    \right)
    \label{eq:Pmatrix}
\end{equation}
with $x=\sqrt{2\sin\frac{\pi}{7}}/\sqrt[4]{21}$, $y=\sqrt{2\sin\frac{2\pi}{7}}/\sqrt[4]{21}$, and $z=\sqrt{2\sin\frac{3\pi}{7}}/\sqrt[4]{21}$. For simplicity, we denote $(3,\pm)$ as $3\pm$. Besides the above nine, there are three more boundary states:
\begin{equation}
  \ket{(s,2)} = \ket{(7-s,2)} 
  =\sum_{s'=1}^3Q_{ss'}\left(\ket{1,s'}-\ket{5,s'}\right)
  \label{eq:symmetricStates}
\end{equation}
where $s=1,2,3$ and
\begin{equation}
  Q_{ss'} = \sqrt{3}\left(
    \begin{array}{ccc}
      x & -y & z \\
      \frac{y^2}{x} & \frac{z^2}{y} & \frac{x^2}{z} \\
      \frac{z^2}{x} & -\frac{x^2}{y} & -\frac{y^2}{z}
    \end{array}
    \right).
    \label{eq:Qmatrix}
\end{equation}

One observation about \eq{eq:symBroken1} and \eq{eq:symBrokenPM} is that the three conformal boundary states $\ket{(s,1)}$, $\ket{(s,3-)}$, and $\ket{(s,3+)}$ can be related to each other by the $Z_3$ transformation, which suggests they represent the $Z_3$ symmetry breaking b.c.'s on the lattice, such as the ordered b.c. with a specific spin. This is because the Ishibashi states $\ket{3\pm,s'}$ correspond to the primary operators with the non-zero $Z_3$ charge discussed in Sec.~\ref{subsec:Z_of_TC3P}: $\phi^{\pm}_{\frac{4}{3}}$ ($s'=1$), $\phi^{\pm}_{\frac{10}{21}}$ ($s'=2$), and $\phi^{\pm}_{\frac{1}{21}}$ ($s'=3$), respectively. Under the $Z_3$ transformation, the Ishibashi states corresponding to those primary operators transform as
\begin{equation}
  \ket{3\pm,s'} \rightarrow \omega^{\pm}\ket{3\pm,s'},
\end{equation}
which relates the conformal boundary states $\ket{(s,1)}$ and $\ket{(s,3\pm)}$ to one another. In fact, the triality, or the invariance of the Dynkin diagram $D_4$ under the $Z_3$ permutation within the three nodes $1$, $(3,-)$, and $(3,+)$, also suggests this relation~\cite{Ruelle1999}.

While the nine boundary states in \eq{eq:symBroken1} and \eq{eq:symBrokenPM} is not invariant under the $Z_3$ transformation, the other three states in \eq{eq:symmetricStates} preserves the $Z_3$ symmetry, since they include no Ishibashi state with non-zero $Z_3$ charge (i.e., $\ket{3\pm,s'}$). Therefore, it is expected that they correspond to the $Z_3$ symmetric b.c.'s on the lattice such as the free b.c.

\subsection{Property of the boundary fixed points}\label{subsec:propFixedPoints}

Now that we have the twelve conformal boundary states of the TC3P model, let us calculate the operator contents for each boundary fixed point and the $g$-values, which help us to grasp the boundary renormalization group (RG) picture~\cite{Affleck1991}.

For the boundary states obtained in Sec.~\ref{subsec:boundaryStates}, calculating the operator contents is a simple task as explained in Ref.~\cite{Behrend1998,Behrend2000}. Consider the finite cylinder where the given two conformal boundary states, $\bra{(s_1,a_1)}$ and $\ket{(s_2,a_2)}$, are assigned to either edge, respectively. For the general minimal models classified as $(G,A_{h-1})$, the partition function $Z_{(s_1,a_1)|(s_2,a_2)}$ placed on such a geometry can be calculated as
\begin{equation}
  Z_{(s_1,a_1)|(s_2,a_2)} = \sum_{\substack{1\leq r\leq g-1\\ 1\leq s\leq h-1}}{N_{ss_1}}^{s_2}{V_{r{a_1}}}^{a_2}\chi_{r,s},
  \label{eq:ZBCFTcalculate}
\end{equation}
where $N_s$ is the fusion matrix given by the Verlinde formula~\cite{Verlinde1988} with the modular matrix $S^{(h)}$:
\begin{equation}
  {N_{ss_1}}^{s_2} = \sum_{1\leq \sigma \leq h-1}\frac{S_{s\sigma}^{(h)}S_{s_1\sigma}^{(h)}{S^{(h)}_{\sigma s_2}}^*}{S_{1\sigma}^{(h)}}.
\end{equation}
The $n\times n$ matrices $V_r$ are so-called fused adjacency matrix of the Dynkin diagram $G$ defined recursively as
\begin{align}
  V_1 &= I, & V_2 &= G, & V_{i+1} = V_2V_i-V_{i-1},
\end{align}
where $I$ represents an identity matrix and there is an abuse of notation: $G$ also represents the adjacency matrix of the Dynkin diagram $G$.

Let us apply \eq{eq:ZBCFTcalculate} to the TC3P BCFT. The fusion matrix $N$ can be calculated using $S^{(h)}$ with $h=7$ as
\begin{align}
  N_1 &= I, &
  N_2 &= \left(
  \begin{array}{cccccc}
    0 & 1 & 0 & 0 & 0 & 0 \\
    1 & 0 & 1 & 0 & 0 & 0 \\
    0 & 1 & 0 & 1 & 0 & 0 \\
    0 & 0 & 1 & 0 & 1 & 0 \\
    0 & 0 & 0 & 1 & 0 & 1 \\
    0 & 0 & 0 & 0 & 1 & 0
  \end{array}
  \right), &
  N_3 &= \left(
  \begin{array}{cccccc}
    0 & 0 & 1 & 0 & 0 & 0 \\
    0 & 1 & 0 & 1 & 0 & 0 \\
    1 & 0 & 1 & 0 & 1 & 0 \\
    0 & 1 & 0 & 1 & 0 & 1 \\
    0 & 0 & 1 & 0 & 1 & 0 \\
    0 & 0 & 0 & 1 & 0 & 0
  \end{array}
  \right),\\
  N_4 &= \left(
  \begin{array}{cccccc}
    0 & 0 & 0 & 1 & 0 & 0 \\
    0 & 0 & 1 & 0 & 1 & 0 \\
    0 & 1 & 0 & 1 & 0 & 1 \\
    1 & 0 & 1 & 0 & 1 & 0 \\
    0 & 1 & 0 & 1 & 0 & 0 \\
    0 & 0 & 1 & 0 & 0 & 0
  \end{array}
  \right), &
  N_5 &= \left(
  \begin{array}{cccccc}
    0 & 0 & 0 & 0 & 1 & 0 \\
    0 & 0 & 0 & 1 & 0 & 1 \\
    0 & 0 & 1 & 0 & 1 & 0 \\
    0 & 1 & 0 & 1 & 0 & 0 \\
    1 & 0 & 1 & 0 & 0 & 0 \\
    0 & 1 & 0 & 0 & 0 & 0
  \end{array}
  \right), &
  N_6 &= \left(
  \begin{array}{cccccc}
    0 & 0 & 0 & 0 & 0 & 1 \\
    0 & 0 & 0 & 0 & 1 & 0 \\
    0 & 0 & 0 & 1 & 0 & 0 \\
    0 & 0 & 1 & 0 & 0 & 0 \\
    0 & 1 & 0 & 0 & 0 & 0 \\
    1 & 0 & 0 & 0 & 0 & 0
  \end{array}
  \right).
\end{align}
The fused adjacency matrices are given as follows:
\begin{align}
  V_1 &= V_5 = I, &
  V_2 &= V_4 = G = \left(
    \begin{array}{cccc}
      0 & 1 & 0 & 0 \\
      1 & 0 & 1 & 1 \\
      0 & 1 & 0 & 0 \\
      0 & 1 & 0 & 0
    \end{array}
    \right), &
  V_3 &= \left(
    \begin{array}{cccc}
      0 & 0 & 1 & 1 \\
      0 & 2 & 0 & 0 \\
      1 & 0 & 0 & 1 \\
      1 & 0 & 1 & 0
    \end{array}
    \right).
\end{align}

Using \eq{eq:ZBCFTcalculate}, we calculate the operator contents for all the boundary fixed points considered in this paper, which is summarized in Tab.~\ref{tab:fps}. For later discussion, we would like to comment on the case with the `degenerated boundary state', i.e., the b.c. specified by the linear combination of the conformal boundary states. Suppose a finite cylinder where on one edge the b.c. $\sum_i\ket{(s_1^i,a_1^i)}$ is imposed while on the other $\sum_j\ket{(s_2^j,a_2^j)}$. On this cylinder geometry, the partition function is just the summation of those with the elementary conformal boundary states, as
\begin{equation}
  Z_{\sum_i(s^i_1,a^i_1)|\sum_j(s^j_2,a^j_2)} = \sum_{i,j}Z_{(s^i_1,a^i_1)|(s^j_2,a^j_2)} = \sum_{\substack{1\leq r\leq g-1\\ 1\leq s\leq h-1}}\left[\sum_{i,j}{N_{ss^i_1}}^{s^j_2}{V_{ra^i_1}}^{a^j_2}\right]\chi_{r,s}.
  \label{eq:ZBCFTcalculateDegenerated}
\end{equation}
\begingroup
\begin{table}
  \caption{The boundary fixed points of the TC3P model.}\label{tab:fps}
  \centering
\renewcommand{\arraystretch}{1.5}
  \begin{tabular}{|c|cc|}\hline
   \begin{tabular}{c}conformal\\boundary state\end{tabular} & partition function & $g$-value \\[2pt]\hline\hline
   $\ket{(1,1)}$, $\ket{(1,3\pm)}$ & $\chi_{1,1}+\chi_{5,1}$ & $x$ \\[4pt]\hline
   $\ket{(2,1)}$, $\ket{(2,3\pm)}$ & $\chi_{1,1}+\chi_{5,1}+\chi_{1,3}+\chi_{5,3}$ & $\frac{y^2}{x}$ \\[4pt]\hline
   $\ket{(3,1)}$, $\ket{(3,\pm)}$ & $\chi_{1,1}+\chi_{5,1}+\chi_{1,3}+\chi_{5,3}+\chi_{1,5}+\chi_{5,5}$ & $\frac{z^2}{x}$ \\[4pt]\hline\hline
   $\ket{(1,2)}$ & $\chi_{1,1}+\chi_{5,1}+2\chi_{3,1}$ & $\sqrt{3}x$ \\[4pt]\hline
   $\ket{(2,2)}$ & $\chi_{1,1}+\chi_{5,1}+\chi_{1,3}+\chi_{5,3}+2\left(\chi_{3,1}+\chi_{3,3}\right)$ & $\frac{\sqrt{3}y^2}{x}$ \\[4pt]\hline
   $\ket{(3,2)}$ & \begin{tabular}{c}$\chi_{1,1}+\chi_{5,1}+\chi_{1,3}+\chi_{5,3}+\chi_{1,5}+\chi_{5,5}$\\$+2\left(\chi_{3,1}+\chi_{3,3}+\chi_{3,5}\right)$\end{tabular} & $\frac{\sqrt{3}z^2}{x}$ \\[15pt]\hline
  \end{tabular}
\end{table}
\endgroup

We also calculate the $g$-values for each fixed point, which represents `the degeneracy of the ground states', defined for the boundary fixed point of $\ket{(s,a)}$ as~\cite{Behrend2000}
\begin{equation}
  g_{(s,a)} = \frac{\braket{1,1|(s,a)}}{\sqrt{S_{11}^{(h)}S_{11}^{(g)}}}
  = \left\{
  \begin{array}{ll}
    P_{s1} & \ (a=1,3\pm) \\
    Q_{s1} & \ (a=2)
  \end{array}
  \right..
  \label{eq:gvalues}
\end{equation}
The results of our calculation are concluded in Tab.~\ref{tab:fps}. Since the $g$-theorem guarantees that the $g$-value decreases along the boundary RG flow from a UV to an IR fixed point~\cite{Affleck1991}, these quantities are useful to determine the boundary phase diagram in the later section.

\section{Lattice realization of the conformal boundary states}\label{sec:lattice}

Now that we have the twelve conformal boundary states which are expected to appear in the TC3P model on a lattice, our next task is to analyze the surface phase transitions of the lattice model with respect to the TC3P BCFT. The goal of this section is to reveal the correspondence between the conformal boundary states found in the BCFT and the physical b.c.'s realized on the lattice. To achieve this, the 3-state dilute Potts model in \eq{eq:S3BC} is studied numerically by the TNR method, which allows us to access the accurate conformal data~\cite{Evenbly2015,Evenbly2016,Evenbly2017}, and also can be applied in the presence of open boundaries~\cite{Iino2020}. First, we determine the accurate location of the bulk tricritical point in \eq{eq:S3BC}, so that the computed conformal data are consistent with the ones exactly known. After locating the tricriticality $\Kcbulk$ and $\Dcbulk$, we study the surface phase diagram spanned by the surface parameters $\Ks$ and $\hs$, and discover the seven different boundary fixed points. By comparing the numerically obtained conformal spectrum and the operator contents of the boundary fixed points for each conformal boundary state, the physical realization of them is discussed.

\subsection{Numerical methods}\label{subsec:method}

In order to extract the conformal spectrum from a Hamiltonian on the lattice, we adopt the TNR algorithm. In this section, we present a brief review of the RG technique in the tensor network formalism and the way of computing the scaling dimensions.

The universal part of the partition function of the critical classical system on a $N\times M$ torus can be, in general, represented as $Z_{\mathrm{CFT}} = \mathrm{Tr}\ T^M$ with the transfer matrix
\begin{equation}
  T = \exp\left[-\frac{2\pi}{N}\left(L_0+\bar{L}_0-\frac{c}{12}\right) \right],
  \label{eq:trfmat_t}
\end{equation}
where $L_0$ and $\bar{L}_0$ are the operators in the Cartan subalgebra of \eq{eq:VirasoroAlgebra} and $c$ is the central charge~\cite{Cardy1986_2}. Since the eigenvalues of $L_0+\bar{L}_0$ yield the scaling dimensions, the diagonalization of the transfer matrix \eq{eq:trfmat_t} after an appropriate normalization allows us to extract the conformal spectrum.

One of the simplest ways of constructing the transfer matrix on the lattice is to employ the tensor network formalism. The transfer matrix \eq{eq:trfmat_t} on the square lattice with a periodic b.c. consists of the rank-4 tensors as
\begin{equation}
  T \propto
  \begin{minipage}{1.1truecm}
    \centering
    \includegraphics[width=2truecm,clip]{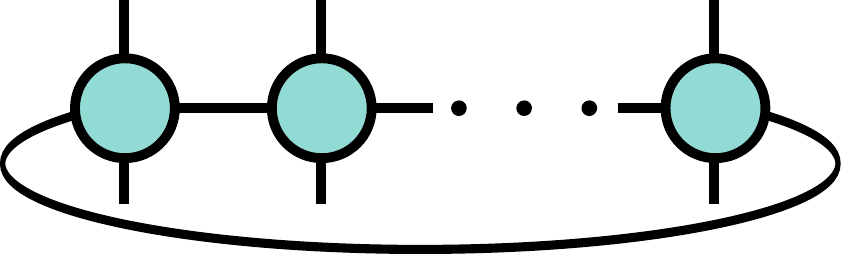}
  \end{minipage}
  \label{eq:trfmat_t_tn}
\end{equation}
where $N$ rank-4 tensors, in each of which the degrees of freedom of one lattice site are implemented, are arranged in a row, and the connected bond indicates that the corresponding index shared by the two tensors is contracted. For the detail on how to obtain the local tensor for a given statistical model, for example see the reference~\cite{Zhao2010}.

Therefore, our task is to construct the transfer matrix \eq{eq:trfmat_t_tn} from the local tensors and diagonalize it to compute the scaling dimensions. However, the exact construction and diagonalization of the transfer matrix requires exponentially exploding numerical cost in terms of the lattice sites. Since it is significant to deal with larger transfer matrices for the purpose of achieving accurate conformal spectrum, more sophisticated methods are essential.

One remedy is to coarse-grain the large transfer matrix by means of some approximation, which can be executed very efficiently in the formalism of the tensor network by the methods called \textit{tensor renormalization group} (TRG)~\cite{Levin2007}. The essence of the TRG methods is to approximate a cluster of tensors as the fewer number of them without increasing the degrees of freedom in the local tensor, based on the philosophy of the real-space RG~\cite{Kadanoff1975}. For example, by replacing the plaquette formed by the four tensors with a single tensor,
\begin{equation}
  \begin{minipage}{1.1truecm}
    \centering
    \includegraphics[width=1.2truecm,clip]{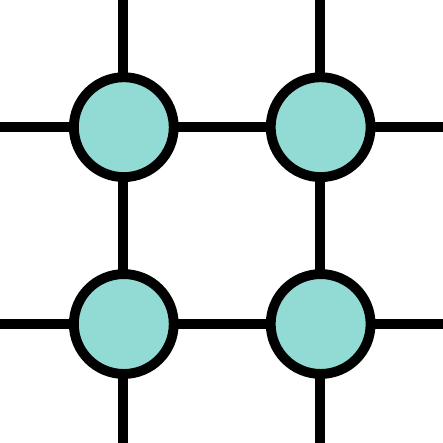}
  \end{minipage}
  \quad=
  \begin{minipage}{1.1truecm}
    \centering
    \includegraphics[width=0.6truecm,clip]{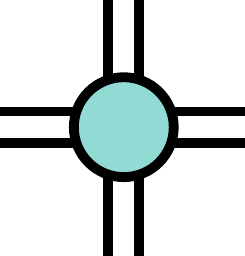}
  \end{minipage}
  \approx
  \begin{minipage}{1.1truecm}
    \centering
    \includegraphics[width=0.6truecm,clip]{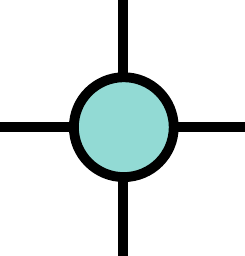}
  \end{minipage},
  \label{eq:trg}
\end{equation}
one can obtain the coarse-grained local tensor where the degrees of freedom of the four original tensors are approximately included. Notice that the point of \eq{eq:trg} is that some threshold $\chi$ should be set for the \textit{bond dimension}, the degree of freedom each bond carries, to avoid the exponential divergence of it. This RG procedure enables us to construct the effectively larger transfer matrix without the exponentially expensive cost, because beginning with an initial tensor representing one lattice site, after $i$ RG steps of \eq{eq:trg} the renormalized local tensor effectively includes the $4^i$ original lattice sites. Although we do not consider in the present paper, we would like to comment that one can easily calculate various physical quantities including the energy and order parameters in the formalism of tensor network with TRG technique.

The TNR method, the algorithm we employ in this work, is also based on the concept in \eq{eq:trg}, which is an improved version of the TRG method so as to simulate even the critical system efficiently. For further explanation and the detailed algorithm, see the references~\cite{Evenbly2015,Evenbly2017}.

In the presence of open boundaries, we can take the same strategy to compute the conformal spectrum accurately~\cite{Iino2019,Iino2020}, where the transfer matrix is expressed as
\begin{align}
  T &= \exp\left[-\frac{\pi}{N}\left(L_0-\frac{c}{24}\right) \right]\\
  \label{eq:trfmat_b}
  & \propto
  \begin{minipage}{1.1truecm}
    \centering
    \includegraphics[width=2truecm,clip]{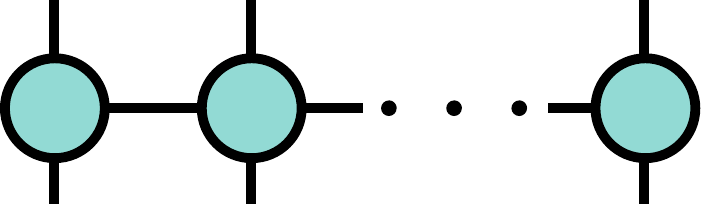}.
  \end{minipage}
\end{align}
The rank-3 tensors in the both sides of the transfer matrix represent the boundaries of the cylinder geometry. Note that we need assume either of the lowest scaling dimension or the central charge by our hand to determine the one we do not assume. In this work, because we only focus on the boundary fixed point, where the same b.c. is imposed on the both boundaries of the cylinder, we assume the lowest scaling dimension $h_0=0$.

Another comment on our numerical method is that the $Z_2$ symmetric tensor is employed, which makes it possible to reduce the computational cost~\cite{Singh2011}. Notice that even in the presence of the boundary external field $\hs$ in \eq{eq:S3BC}, the Hamiltonian possess the $Z_2$ symmetry in terms of the permutation of two spins $B$ and $C$. Also, all the numerical simulations are performed with the bond dimension $\chi=36$.

\subsection{Location of the bulk tricritical point}\label{subsec:location}

\begin{figure}
  \includegraphics[width=1.0\textwidth]{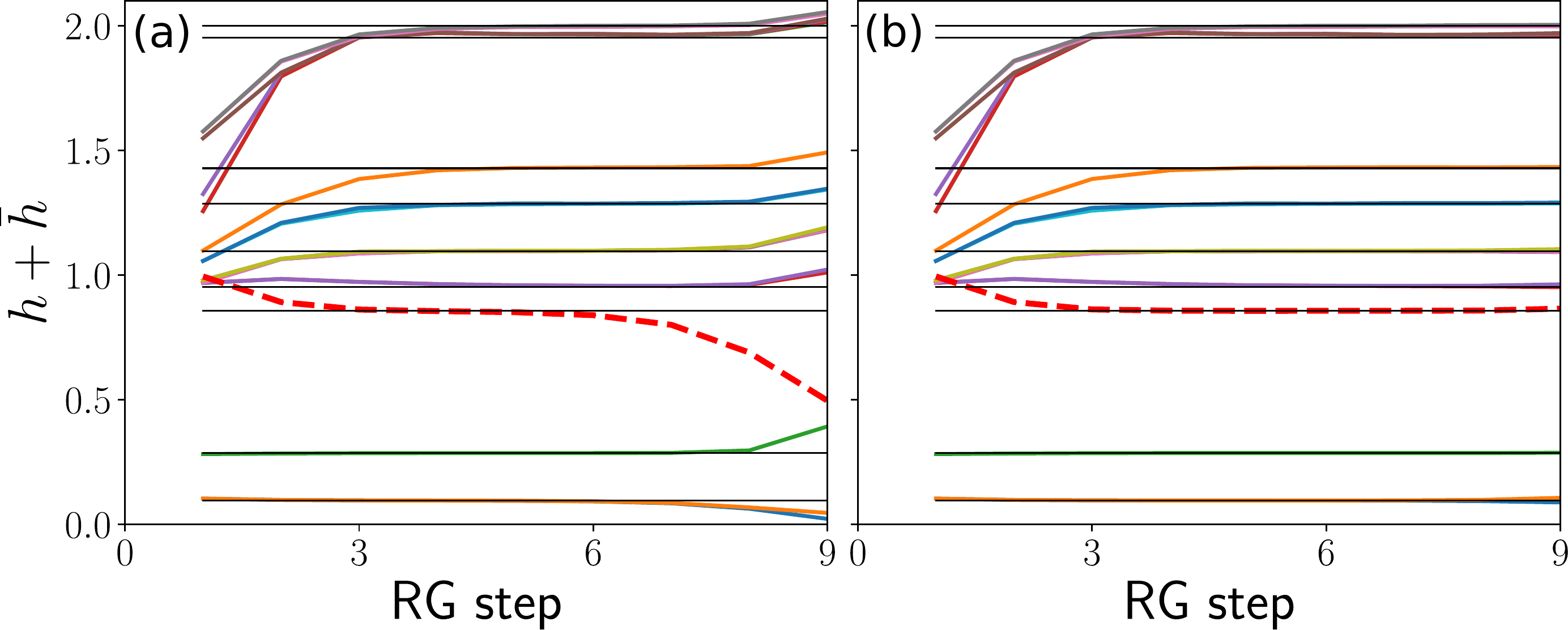}
\caption{(Color online) The computed central charge (red dashed line) and the eighteen lowest scaling dimensions (colored solid lines) for each RG step at (a) $\Kcbulk=1.649913(5)$ and $\Dcbulk=3.152173(10)$, used in the previous study, and (b) $\Kcbulk=1.649850(2)$ and $\Dcbulk=3.152027(1)$ obtained in this study. The black thin solid lines represent the exact values for the TC3P CFT.}
\label{fig:compare}
\end{figure}

Before discussing the surface critical behavior, it is essential to realize the \textit{bulk} tricriticality of the TC3P CFT on the lattice by the fine tuning of the bulk parameters. In the previous study, the bulk tricritical point of \eq{eq:S3BC} is determined by the transfer matrix method, as $\Kcbulk=1.649913(5)$ and $\Dcbulk=3.152173(10)$~\cite{Qian2005}. However, this computed tricritical point turns out to be slightly off critical for our TNR simulation with $\chi=36$, as shown in Fig.~\ref{fig:compare} (a), where the numerically obtained scaling dimensions and central charge are going off the exact values as the RG step (i.e., the system size) grows. Notice that in this simulation we employ the periodic boundary condition and do not consider the surface parameters.

To obtain better tricritical point for our case, we perform the brute-force search in the two-dimensional parameter space spanned by $(K^{\mathrm{bulk}},D^{\mathrm{bulk}})$, where at each pair of the parameters the RG flow of the conformal data is investigated. Our tricritical point is determined so that the computed conformal data at the ninth RG step are as close to the exact values as possible, through which we obtain $\Kcbulk=1.649850(2)$ and $\Dcbulk=3.152027(1)$. In Fig.~\ref{fig:compare} (b) the RG flow computed at the new tricritical point is shown, where the extracted conformal data continue to stay at the exact values stably even for the larger RG steps. In the following discussion on the surface critical behavior, this bulk tricritical point is adopted.

However, notice that we could not conclude that our tricritical point is also more accurate than in Ref.~\cite{Qian2005} in the \textit{thermodynamic limit}, since the extrapolation in terms of $\chi$ is missing in our analysis, while it is performed in terms of the system size in the previous work. We leave it a future work to determine the accurate tricritical points in the $\chi\rightarrow\infty$ limit with tensor network analysis.

\subsection{Surface phase diagram}\label{subsec:SPD}

\begin{figure}
  \centering
  \includegraphics[width=0.6\textwidth]{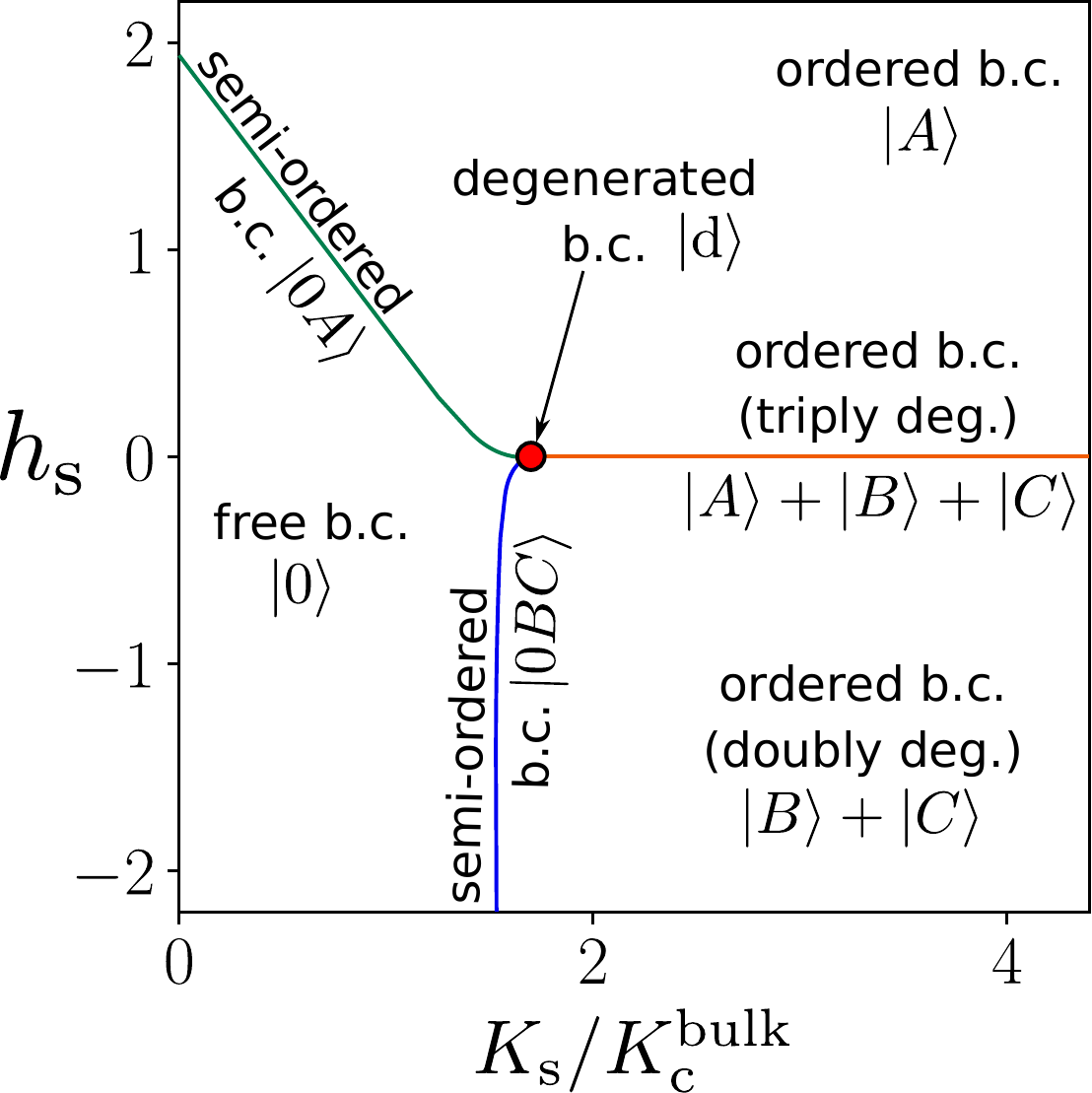}
\caption{(Color online) The schematic surface phase diagram of the 3-state dilute Potts model \eq{eq:S3BC}. The special point of the degenerated b.c. is located at $\Ks/\Kcbulk=1.701(2)$ and $\hs=0$, and the semi-ordered transition line in $\hs>0$ crosses with the $\Ks=0$ line at $h_s=1.938(9)$. On the other hand, the other semi-ordered line in $\hs<0$ has no crossing point with the $\Ks=0$ line, whose end point is at $\Ks/\Kcbulk=1.517(9)$ and $\hs=-\infty$.}
\label{fig:phsdgrm}
\end{figure}

Next, we perform the numerical simulation in the presence of open boundaries and compute the conformal spectrum characterizing the boundary fixed points. For a given pair of the surface parameter $(\Ks,\hs)$ the scaling dimensions are extracted, by which we can determine to what phase the parameter point belongs, since the change of the conformal spectrum allows us to detect the surface phase transitions. In Fig.~\ref{fig:phsdgrm}, we describe the schematic phase diagram with the two surface parameters, where we discover the seven distinct phases. In the following discussion, the naming conventions for the boundary states are based on those in the tricritical Ising model~\cite{Chim1996}.

When $\hs=0$ and $\Ks=0$, the $S_3$ Potts spin is absent on the boundary, which means the edge is occupied by the vacancies, the `$0$' states, and not magnetized~\cite{Deng2004}. We call this boundary state $\ket{0}$, the free b.c., which covers the finite region in the vicinity of the origin.

As is explained in Sec.~\ref{sec:intro}, a remarkable feature of this surface phase diagram is that the edges can be ordered even without the magnetic field, if the surface coupling exceeds some finite threshold. As is also stressed in Sec.~\ref{sec:intro}, this spontaneously symmetry breaking on a one-dimensional edge can be possible owing to the strongly correlated bulk at tricriticality. Such a transition occurs at $\hs=0$ and $\Ks/\Kcbulk=1.701(2)$, which is a special point named as the degenerated b.c., $\ket{\mathrm{d}}$.

Since $\hs>0$ induces the $A$ spin, the boundary can undergo a phase transition into the ordered b.c. with the $A$ spin named as $\ket{A}$, and we call the transition a semi-ordered b.c. labeled as $\ket{0A}$. The phase boundary between the free b.c. phase and the $A$-ordered phase falls into the universality class of this semi-ordered b.c., which crosses with the $\Ks=0$ line at $\hs=1.938(9)$.

On the other hand, the negative $\hs$ suppresses the $A$ spin, which results in the phase transition between the free b.c. phase and another ordered phase with the $B$ or $C$ spin. The boundary state in this phase is labeled as $\ket{B}+\ket{C}$ since the doubly degenerated spectra appears as will be discussed later, while the transition line is again a semi-ordered b.c. labeled as $\ket{0BC}$. Notice that this semi-ordered transition line in $\hs<0$ does not cross with the $\Ks=0$ line, which means the boundary never becomes polarized even for the limit of $\hs\rightarrow -\infty$ if $\Ks$ is sufficiently small. The end point of this transition line at $\hs= -\infty$ is observed at $\Ks/\Kcbulk=1.517(9)$.

The two semi-ordered transition lines merge at the special point. As long as the $S_3$ symmetry is preserved with $\hs=0$, the third ordered phase is stable for the larger $\Ks$ than the special point, whose boundary state is labeled as $\ket{A}+\ket{B}+\ket{C}$ due to the triply degenerated spectra shown in the next subsection.

Finally, we comment that our surface phase diagram in Fig.~\ref{fig:phsdgrm} is qualitatively consistent with the one discussed in the previous study~\cite{Deng2005}.

\subsection{Correspondence to the conformal boundary states in the TC3P BCFT}\label{subsec:CorrCardy}

For the possible boundary fixed points in the BCFT discussed in Sec.~\ref{sec:BCFT}, let us consider to what phases they correspond and understand the phase diagram Fig.~\ref{fig:phsdgrm} with respect to the TC3P BCFT. Because the TNR approach allows us to extract the conformal spectrum from the lattice model, by comparing them with the conformal tower discussed in Sec.~\ref{subsec:propFixedPoints} it can be possible to estimate the correspondence between the conformal boundary states and those realized on the lattice.

\begin{figure}
  \centering
  \includegraphics[width=1.0\textwidth]{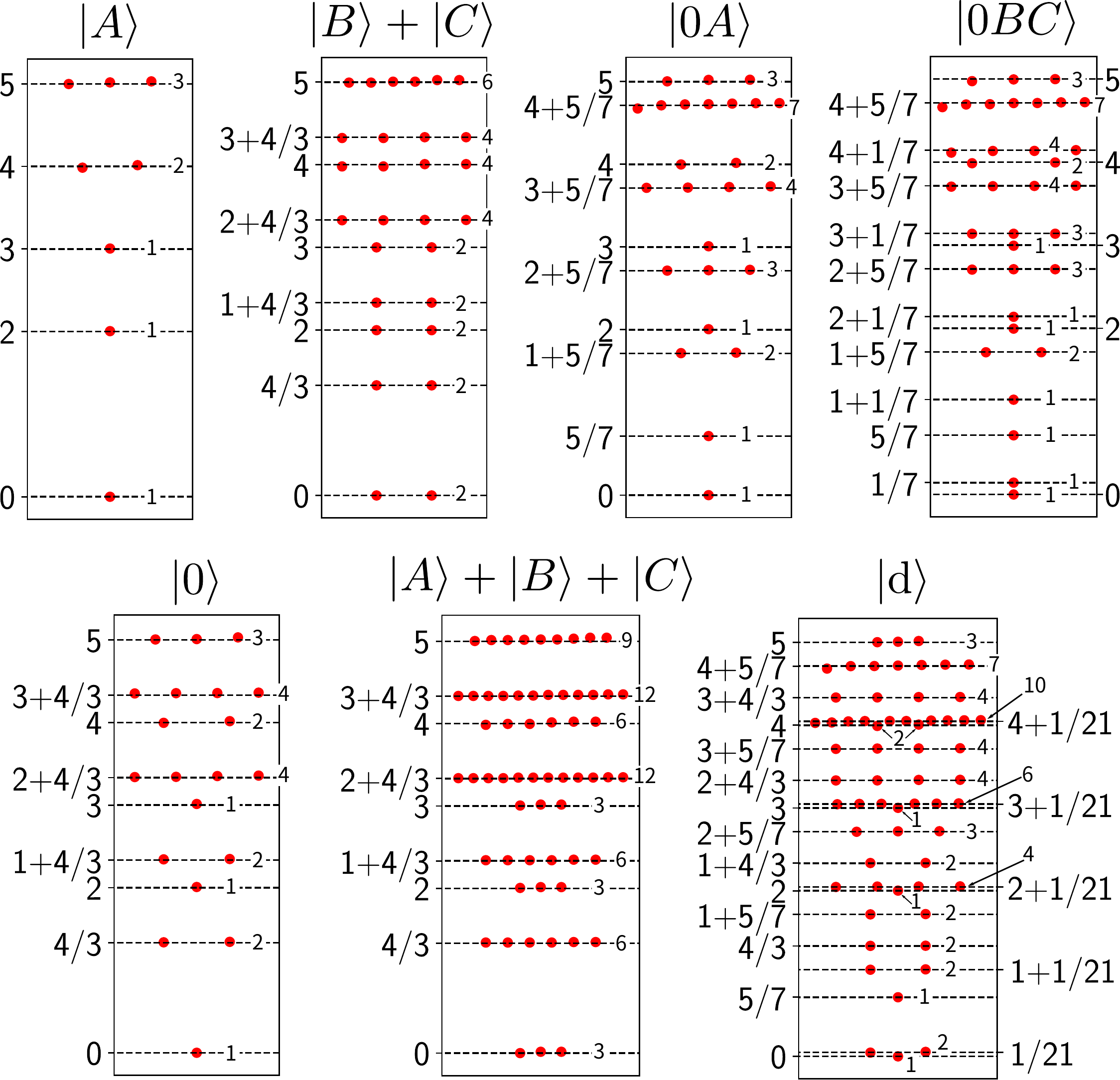}
\caption{(Color online) The numerically extracted lowest scaling dimensions at the characteristic points in every phase in Fig.~\ref{fig:phsdgrm}. The red plots are the numerical results, while the dashed lines and the small figures near the plots are the exactly known scaling dimensions and the degeneracy of them, respectively. The concrete parameters with which the simulations are performed are concluded in Tab.~\ref{tab:conclusion}. All the spectrum are computed at the $8$th RG step.}
\label{fig:spectrum}
\end{figure}

We show the numerical results of the computed spectrum in Fig.~\ref{fig:spectrum} for the seven distinct phases in the surface phase diagram Fig.~\ref{fig:phsdgrm}. The concrete locations of the points where the spectrum are computed, the pairs of the surface parameters $(\Ks,\hs)$, are summarized in Tab.~\ref{tab:conclusion}. For instance, the spectra for the ordered b.c. with the single Potts spin $A$ is extracted at $\Ks=0$ and $\hs=\infty$, whose operator content is consistent with the partition function $\chi_{1,1}+\chi_{5,1}$. This indicates that the conformal boundary state $\ket{(1,1)}$ corresponds to the fixed point of the ordered b.c. $\ket{A}$.
\begingroup
\begin{table}
  \centering
  \caption{The list of the boundary fixed points of the TC3P model observed in the surface phase diagram Fig.~\ref{fig:phsdgrm}. The second column represents the location in the parameter space where the numerical results in Fig.~\ref{fig:spectrum} are computed. The third column shows the estimated operator content from the numerical simulation in Fig.~\ref{fig:spectrum}, from which the corresponding conformal boundary state can be determined as in the fourth column. Notice that $\chi_{h}$ represents the Virasoro character for the primary operator with the conformal weight $h$. Finally, the $g$-values calculated by \eq{eq:gvalues} are denoted in the fifth column.}\label{tab:conclusion}
\renewcommand{\arraystretch}{1.5}
  \begin{tabular}{|cc|ccc|}\hline
    b.c. & $(\Ks/\Kcbulk,\hs)$ & partition function & \begin{tabular}{c}b.c.\\in BCFT\end{tabular} & $g$-value \\[2pt]\hline\hline
    $\ket{A}$ & $(0,\infty)$ & $\chi_0+\chi_5$ & $\ket{(1,1)}$ & $x\approx 0.44$ \\[4pt]\hline
    $\ket{B}+\ket{C}$ & $(\infty,-\infty)$ & $2\left(\chi_0+\chi_5+\chi_{\frac{4}{3}}\right)$ & $\ket{(1,{\boldsymbol 3})}$ & $2x\approx 0.87$ \\[4pt]\hline
    $\ket{0A}$ & $(0,1.9389)$ & $\chi_0+\chi_5+\chi_{\frac{5}{7}}+\chi_{\frac{12}{7}}$ & $\ket{(2,1)}$ & $\frac{y^2}{x}\approx 0.78$ \\[4pt]\hline
    $\ket{0BC}$ & $(1.5179,-\infty)$ & \begin{tabular}{c}$\chi_0+\chi_5+\chi_{\frac{1}{7}}+\chi_{\frac{22}{7}}$\\$+\chi_{\frac{5}{7}}+\chi_{\frac{12}{7}}$\end{tabular} & $\ket{(3,1)}$ & $\frac{z^2}{x}\approx 0.98$ \\[4pt]\hline\hline
    $\ket{0}$ & $(0,0)$ & $\chi_0+\chi_5+2\chi_{\frac{4}{3}}$ & $\ket{(1,2)}$ & $\sqrt{3}x\approx 0.75$ \\[4pt]\hline
    $\ket{\mathrm{d}}$ & $(1.7012,0)$ & \begin{tabular}{c}$\chi_0+\chi_5+\chi_{\frac{5}{7}}+\chi_{\frac{12}{7}}$\\$+2\left(\chi_{\frac{1}{21}}+\chi_{\frac{4}{3}}\right)$\end{tabular} & $\ket{(2,2)}$ & $\frac{\sqrt{3}y^2}{x}\approx 1.36$ \\[13pt]\hline
    \begin{tabular}{r}$\ket{A}+\ket{B}$\\$+\ket{C}$\end{tabular} & $(\infty,0)$ & $3\left(\chi_0+\chi_5+2\chi_{\frac{4}{3}}\right)$ & \begin{tabular}{c}$\ket{(1,1)}$\\$+\ket{(1,{\boldsymbol 3})}$\end{tabular} & $3x\approx 1.31$ \\[4pt]\hline
  \end{tabular}
\end{table}
\endgroup

Because the three conformal boundary states $\ket{(1,1)}$ and $\ket{(1,3\pm)}$ are associated by the $Z_3$ transformation as is explained in Sec.~\ref{subsec:boundaryStates}, we can conclude that the conformal boundary states $\ket{(1,3\pm)}$ correspond to $\ket{B}$ and $\ket{C}$, respectively. This can be verified through the fact that the spectra for the boundary state $\ket{B}+\ket{C}$ is consistent with that of the boundary state $\ket{(1,3-)}+\ket{(1,3+)}\equiv\ket{(1,{\boldsymbol 3})}$. Notice that the partition function $Z_{(1,{\boldsymbol 3})|(1,{\boldsymbol 3})}$ can be calculated from \eq{eq:ZBCFTcalculateDegenerated}, as
\begin{equation}
    Z_{(1,{\boldsymbol 3})|(1,{\boldsymbol 3})} = \sum_{\substack{a_1=3\pm\\a_2=3\pm}}Z_{(1,a_1)|(1,a_2)} = 2\left(\chi_{1,1}+\chi_{5,1}+\chi_{3,1}\right).
\end{equation}
The coincidence of the spectra for $\ket{A}+\ket{B}+\ket{C}$ with that of $\ket{(1,1)}+\ket{(1,{\boldsymbol 3})}$ also supports the above statement.

For the other $Z_3$ symmetry breaking conformal boundary states, the corresponding lattice realization can be found based on the consistency of the spectrum as shown in Tab.~\ref{tab:conclusion}. Namely,
\begin{align}
  \ket{0A} &= \ket{(2,1)}, & \ket{0BC} &= \ket{(3,1)},
\end{align}
which leads to the correspondence of the other Cardy states related by the $Z_3$ transformation:
\begin{align}
  \ket{0B} &= \ket{(2,3+)}, & \ket{0C} &= \ket{(2,3-)},\\
  \ket{0CA} &= \ket{(3,3+)}, & \ket{0AB} &= \ket{(3,3-)}.
\end{align}

Since the boundary states $\ket{0}$ and $\ket{\mathrm{d}}$ possess the $Z_3$ symmetry because of $\hs=0$ and no degenerated ground state in their conformal spectrum, it would be natural to relate them to the boundary states in \eq{eq:symmetricStates}. Judging from the conformal spectrum, we can see that
\begin{align}
  \ket{0} &= \ket{(1,2)}, & \ket{\mathrm{d}} &= \ket{(2,2)}.
\end{align}

Now we are able to identify the lattice realization for the eleven conformal boundary states as concluded in Tab.~\ref{tab:conclusion}, while the physical picture of only $\ket{(3,2)}$ remains unidentified. A remarkable characteristics of $\ket{(3,2)}$ is that in the operator content of its boundary fixed point all the primary operators in the TC3P model appear (see Tab.~\ref{tab:fps}), which means the fixed point is quite unstable with the five relevant scaling fields. As far as we search the two-parameter surface phase diagram, such a fixed point cannot be discovered with $\hs\in\mathbb{R}$ and $\Ks\geq 0$.

A similar conformal boundary state is also found in the 3-state Potts BCFT, whose boundary fixed point contains all the possible primary operators in that BCFT~\cite{Affleck1998,Fuchs1998}. It is revealed that such a boundary state, named as `new' b.c., can be realized in the quantum 3-state Potts chain with an imaginary magnetic boundary field or in the classical two-dimensional Potts model with a boundary negative Boltzmann weight, which can never be accessible with the `physically sound' surface coupling and magnetic field~\cite{Behrend2001}.

In fact, Ruelle conjectured that the TC3P BCFT would possess a similar boundary state to the new b.c. in the 3-state Potts BCFT~\cite{Ruelle1999}. Judging from the numerical investigation, we also guess that $\ket{(3,2)}$ would exist out of the surface phase diagram in Fig.~\ref{fig:phsdgrm} and might be realized with some nonphysical Boltzmann weight. Determination of the boundary Boltzmann weight for this state would be more difficult than the case of the 3-state Potts BCFT, because the duality analysis of the lattice Hamiltonian~\cite{Affleck1998} is more complicated. Then, we also name $\ket{(3,2)}$ as the `new' b.c. and leave it an open problem to consider its realization on the lattice.

Finally, we confirm the surface phase diagram in Fig.~\ref{fig:phsdgrm} in terms of the $g$-values, which are concluded in Tab.~\ref{tab:conclusion}. For the phase transition between the free b.c. and the $A$-ordered phase, an inequality
\begin{equation}
  g_A < g_0 < g_{0A}
\end{equation}
indicates the fixed point of $\ket{0A}$ is more unstable than that of $\ket{A}$ and $\ket{0}$, which is consistent with the RG picture in Fig.~\ref{fig:phsdgrm}. Similarly,
\begin{align}
  &g_0 < g_{B+C} < g_{0BC} & &\mathrm{and} & &g_A < g_{B+C} < g_{A+B+C}
\end{align}
are consistent with the two phase transitions of $\ket{0}\leftrightarrow\ket{B}+\ket{C}$ and $\ket{A}\leftrightarrow\ket{B}+\ket{C}$. Furthermore, the $g$-value of the special point satisfies
\begin{equation}
  g_{0A} < g_{0BC} < g_{A+B+C} < g_{\mathrm{d}},
\end{equation}
which is also consistent with the phase diagram. We comment that the new b.c. is the most unstable fixed point with $g_{\mathrm{new}}=\frac{\sqrt{3}z^2}{x}>g_{\mathrm{d}}$ in those appearing in the phase diagram Fig.~\ref{fig:phsdgrm}.

\section{Conclusion and discussion}\label{sec:conclusion}

In this paper, we investigate the surface critical behavior of the TC3P model in two dimension more precisely than in the previous work with the MC simulation. We first review the BCFT for the TC3P universality class, where we make a list of the conformally invariant b.c.'s with the $g$-values and the operator contents for the possible boundary fixed points, based on the $A$-$D$-$E$ classification provided in Ref.~\cite{Behrend2000}. The correspondence between the obtained boundary fixed points and the phase diagram of the lattice model is studied with the tensor network method, by which the accurate conformal data emergent on the lattices are numerically accessible. After determining the location of the bulk tricritical point in the 3-state dilute Potts model, we study the surface phase diagram spanned by the surface couplings and magnetic fields. Our surface phase diagram is qualitatively consistent with the one in the previous study, where the seven distinct phases are discovered. Comparing the numerically extracted conformal spectrum with that of each boundary fixed point calculated exactly, we are able to identify the lattice realizations of the conformal boundary states in the TC3P BCFT except for the one named as `new' b.c., which might be realized on the lattices with physically unsound Boltzmann weight similarly to the new b.c. in the 3-state Potts BCFT.

The most straightforward way of calculating the boundary weight for the new b.c. would be to make use of the critical $A$-$D$-$E$ \textit{dilute} lattice models~\cite{Warnaar1992,OBrien1995}. The calculation of the boundary weights for this dilute series can be performed with the dilute Temperley-Lieb algebra using the same strategy discussed by Behrend and Pearce~\cite{Behrend2001}, although the problem is, as suggested in the discussion of their paper, the analysis of the dilute lattice models are more complicated than the ordinary ones since the dilute Temperley-Lieb algebra includes more generators than the ordinary Temperley-Lieb one.

It is, in general, significant to verify whether the CFTs are consistent with the critical phenomena in actual lattice models, since the emergent conformal symmetry in critical systems is only a conjecture except very few cases exactly studied. Particularly in the presence of open boundaries, owing to a rich variety of conformal b.c.'s occurring in BCFTs, studying the lattice realization of them for various lattice models is an intriguing problem. We would like to stress that for such a purpose our programs employed in this paper are easily applicable for other two-dimensional classical or one-dimensional quantum systems at criticality.

It would also be interesting to investigate the more general multicritical Ising universality classes, which would correspond to the minimal CFTs with larger central charges. Since the number of the primary fields grows as the central charge becomes closer to $c=1$, it is expected that the higher-multicritical models have more conformal boundary states, which would suggest the richer surface phase diagram. On the other hand, the higher multicritical 3-state Potts model could not be described by the minimal CFTs~\cite{Zamolodchikov1987}, which suggests that it would be difficult to calculate the complete list of the conformal b.c.'s in the framework of BCFT. We would like to suggest, even in such a case, construction and numerical study of the lattice model is possible, which might be useful to discover unfound conformal boundary states in the BCFT.

\appendix

\begin{acknowledgements}
  S.I. thanks Paul A Pearce, J{\"u}rgen Fuchs, and Naoki Kawashima for the useful comments, and Satoshi Morita for providing the code of the $Z_q$ symmetric tensor. He is also grateful to one of the anonymous referees for pointing out Ref.~\cite{Behrend2000} and the incorrect discussion in the former version of this paper. Finally, he thanks the support of Program for Leading Graduate Schools (ALPS).
\end{acknowledgements}

%
%




\end{document}